\documentclass[12pt]{article}
\usepackage{graphicx} 
\newcommand{\ra}{\rightarrow}

\newcommand{\nit}{\noindent}
\newcommand{\no}{\nonumber}
\newcommand{\be}{\begin{equation}}
\newcommand{\ee}{\end{equation}}
\newcommand{\ba}{\begin{eqnarray}}
\newcommand{\ea}{\end{eqnarray}}

\newcommand{\dta}{\mbox{$\delta$}}

\newcommand{\al}{\mbox{$\alpha$}}

\newcommand{\eps}{\mbox{$\epsilon$}}
\newcommand{\om}{\mbox{$\omega $}}
\newcommand{\gam}{\mbox{$\gamma$}}
\newtheorem{theo}{Theorem}[section]

\newtheorem{lem}{Lemma}[section]

\newtheorem{rmk}{Remark}[section]

\begin{document}
\thispagestyle{empty}
\title{Global well-posedness and multi-tone solutions 
of a class of nonlinear nonlocal cochlear models in hearing} 
\author{Jack Xin\thanks{Department of Mathematics and TICAM,
University of Texas at Austin,
Austin, TX 78712, USA. Corresponding author, email: jxin@math.utexas.edu.
This work was partially supported by ARO grant DAAD 19-00-1-0524 
and NSF ITR-0219004.} \hspace{.05 in} 
and 
\hspace{.05 in}Yingyong Qi \thanks{
Qualcomm Inc, 5775 Morehouse Drive, San Diego, CA 92121, USA.}
}
\date{}
\maketitle
\baselineskip=18pt

\begin{abstract}
We study a class of nonlinear nonlocal cochlear models of 
the transmission line type, describing the motion of 
basilar membrane (BM) in the cochlea. They are damped dispersive 
partial differential equations (PDEs) driven by time dependent boundary 
forcing due to the input sounds. The global well-posedness in time 
follows from energy estimates. Uniform bounds of solutions hold in
case of bounded nonlinear damping. When the input sounds are multi-frequency 
tones, and the nonlinearity in the PDEs is cubic, 
we construct smooth quasi-periodic solutions (multi-tone solutions) 
in the weakly nonlinear regime, where new frequencies are generated 
due to nonlinear interaction. When the input is two tones at 
frequencies $f_1$, $f_2$ ($f_1 < f_2$), and high enough intensities, 
numerical results illustrate the formation of combination tones at 
$2 f_1 -f_2$ and $2f_2 -f_1$, in agreement with hearing experiments. 
We visualize the frequency content of solutions through  
the FFT power spectral density of displacement at selected  
spatial locations on BM.

\end{abstract}
\thispagestyle{empty}

\newpage

\section{Introduction}
\setcounter{equation}{0}
\setcounter{page}{1}
Digital signal processing on sounds is an essential component  
of modern hearing devices \cite{Pohl}, and a useful tool  
for evaluating acoustic theories of peripheral auditory systems,
\cite{Meddis_01} among others. 
A fundamental issue is to model the auditory response to complex tones
because the nonlinear interaction of 
acoustic waves of different frequencies allows for 
audio compression \cite{Pohl} among other applications.
Nonlinearities are known to originate 
in the cochlea and are further modified in higher level auditory pathways. 
The cochlear mechanics has first principle 
descriptions, and so partial differential equations (PDEs) 
become a natural mathematical framework to initiate computation.   
However, in vivo cochlear dynamics is not a pure mechanical problem, 
and neural couplings are present to modify responses. 
To incorporate both aspects, 
a first-principle based PDE model was studied in \cite{Xin_02} 
for voice signal 
processing, where the neural aspect is introduced in the model 
phenomenologically.  
The first principle based PDE approach 
is more systematic compared with filter bank method \cite{Meddis_01},  
and has shown encouraging results.
In \cite{Xin_02}, time domain computation on multi-tone inputs 
revealed tonal suppressions in qualitative agreement with 
earlier neural experimental findings.

In this paper, we shall analyze the well-posedness and construct  
multitone solutions of such PDE models in the form:
\ba
 & & p_{xx} - N u_{tt} = \epsilon_{s}(x) u_{t}, \;\; x \in (0,L), \label{C0} 
  \\ & & p= m u_{tt} + r(x,|u|,|u_t|) u_{t} + s(x) u, \label{C1}
\ea
where $p$ is the fluid pressure difference across the basilar membrane (BM),
$u$ the BM displacement, $L$ the longitudinal length of BM;
$N$ a constant depending on fluid density and cochlear
channel size; $\epsilon_s (x) \geq 0$ is
the damping of longitudinal fluid motion;
$m$, $r$, $s$ are the mass,
damping, and stiffness of BM per unit
area, with $m$ a constant, $s$ a continuously differentiable
nonnegative function of $x$.
The coefficient $r$ is a nonlinear function(al) of $x$, $u$, $u_t$:
\be
r(x,|u|,|u_t|)= r_a(|u_t|^2) +\gam \, \int_{0}^{L}\, P(|u(x',t)|)
K(x -x')\, dx'. \label{C2}
\ee
Here: (H1) $r_a (\cdot)$ is the local part of BM damping, it is
a nonnegative continuously differentiable monotone increasing
function, $r_a (0) > 0$.
In the nonlocal BM
damping: (H2) $K=K(x)$ is a localized Lipschitz continuous
kernel function with total
integral over $x \in R^1$ equal to 1;
(H3) $P(\cdot)$ is a nonnegative continuously
differentiable function such that for some constant $C > 0$:
\be
P(0)=0, P(q) \leq C (1 + q^2),\;\; \forall q \geq 0. \label{GC}
\ee

The boundary and initial conditions of the system are:
\ba
& & p_{x}(0,t)= T_{M} p_{T}(t)\equiv f(t),\;\;  p(L,t)=0, \label{C4} \\
& & u(x,0)=u_0(x), u_{t}(x,0) = u_1(x), \label{C5}
\ea
where the initial data is such that $(u_0, u_1)(L)=(0,0)$; 
$p_{T}(t)$ is the input sound pressure at the eardrum; and $T_M$ is a
bounded linear map modeling functions of middle ear,
with output depending on the frequency content of $p_T(t)$.
If $p_T = \sum_{j=1}^{J_M} A_j \exp\{i\omega_j t\} + c.c.$,
a multi-tone input, c.c denoting complex conjugate, $J_M$ a positive
integer,
then $T_M p_T(t) =\sum_{j=1}^{J_M} B_j \exp\{i\omega_j t\}+ c.c$, where 
$B_j = a_{M}(\omega_j) A_j$, c.c for complex conjugate,
$a_M (\cdot) $ a scaling function built 
from the filtering characteristics of the
middle ear \cite{Guin_Peake_66}.

Cochlear modeling has had a long history, and various 
linear models have been studied at length by analytical and 
numerical methods, \cite{Keller_85}, \cite{Peskin_88} 
and references therein. A brief derivation of the
cochlear model of the transmission line type, e.g. the  
linear portion of (\ref{C0})-(\ref{C5}), is nicely presented in 
\cite{Sondhi_80} based on fluid and elasticity equations.
 
It has been realized that 
nonlinearity is essential for multitone interactions, 
\cite{Hall_77,Kim_86,Boer_96,Geisler_98} etc. 
Nonlinearity could be introduced 
phenomenologically based on 
spreading of electrical and neural 
activities between hair cells at different BM 
locations suggested by 
experimental data, \cite{Jau_Geisler_83}, \cite{Deng_92}. 
Such a treatment turned out to be efficient for signal processing 
purpose \cite{Xin_02}, 
and (\ref{C2})-(\ref{GC}) is a generalization of existing 
nonlinearities \cite{Jau_Geisler_83}, \cite{Deng_92}, \cite{Strube_85}.

Multitone solutions require one to 
perform numerical computation in the time domain.
The model system (\ref{C0})-(\ref{C5}) is dispersive, and long waves 
tend to propagate with little decay from entrance point $x=0$ (stapes) 
to the exit $x=L $ (helicotrama). 
The function $\eps_s (x)$ is supported near $x=L$, its role in numerics 
is to suck out the long waves accumulating near the exit \cite{Xin_02}.  
Selective positive or negative damping has been  
a novel way to filter images in PDE method of image processing 
\cite{Osher_90}. In analysis of model solutions that concern mainly 
with interior properties however, we shall set $\eps_s $ to zero 
for technical 
convenience. 

The rest of the paper is organized as follows.
In section 2, we perform energy estimates of solutions 
for the model system (\ref{C0})-(\ref{C5}), prove the global well-posedness 
and obtain growth and uniform bounds in Sobolev spaces.  
In section 3, we construct exact multi-frequency solutions 
when $\gamma $ is small enough and nonlinearity is cubic, 
using contraction mapping in a suitable 
Banach space. The constructed solutions contain all 
linear integral combinations 
of input frequencies. In section 4, for two input tones 
with frequencies $f_1$ and $f_2$ ($f_1 < f_2$), we illustrate 
numerically the generated 
combination tones $2 f_1 - f_2 $ and $2 f_2 -f_1$ on 
power spectral density plots at selected points on BM. These tones 
are heard on musical instruments (piano and violin), in particular, 
$2 f_1 - f_2 $ is known as the Tartini tone. 
The conclusions are in section 5.

\section{Global Well-Posedness and Estimates}
\setcounter{equation}{0}
Let us consider the initial boundary value problem (IBVP) formed 
by (\ref{C0})-(\ref{C5}) and show that solutions exist uniquely 
in a proper function space for all time. To this end, 
it is convenient to work with the equivalent integral form of the 
equations. It follows from (\ref{C0}) and (\ref{C4}) that:
\[
p_{x} = \int_{0}^{x}\, ( N u_{tt} +\eps_s (x) u_t )\, dx + f(t), 
\]
\be
-p(t,x)= \int_{x}^{L}\, dx' \, \int_{0}^{x'}\, (N u_{tt} + \eps_s  u_{t}) \,
dx'' + f(t)(L-x). \label{W1}
\ee
Combining (\ref{C1}) and (\ref{W1}), we get:
\be
m\, u_{tt} + \int_{x}^{L}\, dx'\, \int_{0}^{x'}\, (N u_{tt} +\eps_s u_{t})\, 
dx'' + f(t)(L-x) = -r(x,|u|,|u_t|)u_{t} - s(x) u, \label{W2}
\ee
with initial data (\ref{C5}). Let $w = (w_1,w_2) = (u,u_t)$, and write 
(\ref{W2}) into the system form:
\ba
w_{1,t} & = & w_2, \label{W3} \\
m w_{2,t} + \int_{x}^{L}\, dx' \, \int_{0}^{x'}\, N\, w_{2,t}\, dx''
& = & - \int_{x}^{L} \, dx' \, 
\int_{0}^{x'}\, \eps_{s} \, w_{2}\, dx'' \no \\
& & - r(x,|w_1|,|w_2|)\, w_2 - s(x)\, w_1 + f(t)\, (x -L). \label{W4}
\ea
The related integral form is:
\ba
w_1 & = & u_{0} + \int_{0}^{t}\, w_{2}(x,\tau)\, d\tau, \no \\
Aw_2 & = & Au_{1} - \int_{0}^{t}\, d\tau \, \int_{x}^{L}\, dx'\, 
\int_{0}^{x'}\, \eps_s w_2 \, dx'' \no \\
& & -\int_{0}^{t}\, d\tau \, r(x,|w_1|,|w_2|)\, w_2 - s(x)\int_{0}^{t}\, 
w_1\, d\tau + (x-L)\int_{0}^{t}\, f(\tau)\, d\tau, \label{W5}
\ea
where $A: L^{2}([0,L]) \ra L^{2}([0,L])$ is a bounded self-adjoint 
linear operator:
\be
A\, g \equiv m\, g + \int_{x}^{L}\, dx' \, \int_{0}^{x'}\, N \, g \, dx''
\equiv m\,g + \tilde{A}\, g.
\label{W6}
\ee

\nit To see the self-adjointness of $A$, let $g$, $h \in L^2([0,L])$, then 
$(\tilde{A}g,h)=(g,\tilde{A}h)$ or:
\ba
& & \int_{0}^{L}\left ( \int_{x}^{L}\, dx'\, \int_{0}^{x'}\, g(x'') 
\, dx'' \right ) \, h(x)\, dx = \int_{0}^{L}\, \left ( \int_{x}^{L}\, dx' 
\int_{0}^{x'}\, g(x'')\, dx'' \right )\, d \int_{0}^{x} h \no \\
& = & \int_{0}^{L}\, \left ( \int_{0}^{x} \, dx' \, g(x') \right )
\left (  \int_{0}^{x} \, dx' \, h(x') \right )\, dx,  \no
\ea
hence $(A g, h)_{L^2} = (g, Ah)_{L^2}$. Clearly, $A$ is bounded; also
$(A\cdot, \cdot ) = (A\cdot, \cdot )_{L^2}$ is 
an equivalent square $L^2$ norm:
\ba
& & (A g, g) = m \| g\|_{2}^{2} + N 
\int_{0}^{L} (\int_{0}^{x}\, g)^2\, dx \geq m \| g\|_{2}^{2}, \no \\
& & (A g, g) \leq m \| g\|_{2}^{2} + N L^2 \|g\|_{2}^{2} = 
(m + N L^2) \|g\|_{2}^{2}. \label{W7}
\ea
Moreover, $A$ has a bounded inverse. To see this, note that 
$A - m \, Id $ is a compact operator on $L^2([0,L])$, so Riesz-Schauder 
theory \cite{Zeidler_95} says that the spectrum of 
$A$ can have only eigenvalues of 
finite multiplicities except at number $m$. On the other hand, 
zero cannot be an eigenvalue of $A$, as $(A g, g) \geq m \| g\|_{2}^{2}$. 
The bounded inverse of 
$A$ follows, and we denote it by $A^{-1}$ below.

Now we establish the global existence of solutions of 
(\ref{W3})-(\ref{W4}) in the function space $C([0,\infty); (H^{1}([0,L]))^2)$. 
It is straightforward to show by contraction mapping principle that 
if $\|(u_{0},u_{1}) \|_{H^1} < \infty $, 
there is a time $t_*$ such that (\ref{W5}) 
has a unique solution in $C([0,t_*); (H^{1}([0,L]))^2)$ under 
our assumptions on the nonlinearities. Such a solution 
in fact lies in $C^1([0,t_*); (H^{1}([0,L]))^2)$, and obeys the 
differential form of equations (\ref{W3})-(\ref{W4}), with 
both sides interpreted in the $H^1$ sense. Taking the limit $x \to L$, 
we find that the system (\ref{W3})-(\ref{W4}) reduces to the ODE system:
\ba
w_{1,t} & = & w_2, \no \\
w_{2,t} & = & - r(t) \, w_2 - s(L) w_1, \no 
\ea
with initial data $(w_1,w_2)(L,0) = (0,0)$, hence $(w_1,w_2)(L,t) = (0,0)$, 
$\forall t \in (0, t_*)$.

Let us derive global in time 
estimates of solutions in $H^1$ to extend the local solutions to global ones 
(so $t_* = \infty$). 
The left hand side of (\ref{W4}) is just $(A w_2)_t$, 
and:
\[ (w_2,(Aw_2)_t) = (w_2, A(w_2)_t) = (Aw_2, w_{2,t}) = (w_{2,t}, A w_2), \]
so:
\be
{d\over dt} (A w_2, w_2 ) = (w_{2,t}, A w_2) + (A w_{2,t}, w_2) =  
2 (w_2, A w_{2,t}), \label{W8}
\ee
hence ${1\over 2} {d\over dt} (A w_2, w_2) = (w_2, A w_{2,t})$. 
Multiplying (\ref{W3}) by $w_1$, (\ref{W4}) by $w_2$, adding the two 
expressions and integrating over $[0,L]$, we estimate with 
Cauchy-Schwarz inequality:
\ba
& & (w_1, w_{1,t}) + (w_2, (A w_2)_t)  =   {1\over 2}{d\over dt}( (w_1,w_1) + 
(Aw_2, w_2)) \no \\
& = & - \left ( \int_{x}^{L}\, dx'\, \int_{0}^{x'} \eps_{s} w_2\, dx'', w_2 
\right ) - (r \, w_2, w_2)   \no \\
& & + (w_2,w_1) - (s w_1, w_2) + (f(t)\, (x-L), w_2 ) \no \\
& \leq & - r_{a}(0) \|w_2\|^{2}_{2}  + r_{a}(0)\|w_2\|^{2}_{2}  +
{1\over 4 r_{a}(0)} \|f(t)(x-L)\|^{2}_{2} \no \\
& & -\int_{0}^{L} \left ( \int_{0}^{x}\, \eps_s w_2 \right ) 
\left ( \int_{0}^{x} w_2 \right )\, dx + \|(1 -s)\|_{\infty} \|w_2\|_{2} 
\|w_1 \|_{2} \no \\
& \leq &  {1\over 12 r_a(0)} |f(t)|^2 L^3 + L^2 \|\eps_s\|_2 \|w\|_{2}^{2}
+{1\over 2} \|1 -s \|_{\infty} (\|w_1\|_{2}^{2} + \|w_2\|_{2}^{2}) \no \\
& \leq & ({1\over 2} \|1 -s \|_{\infty} + L^2 \|\eps_s\|_{2})\|w_2\|_{2}^{2}
+{1\over 2} \|1 -s \|_{\infty}\|w_1\|_{2}^{2} + 
{1\over 12 r_a(0)} |f(t)|^2 L^3. \label{W9}
\ea
Let $C_1 = \max ({1\over m} \|1 -s \|_{\infty} + {2L^2\over m} \|\eps_s\|_{2}, 
{1\over 2} \|1 -s \|_{\infty})$, and $ 2 E = (w_1,w_1)+(Aw_2,w_2)$, 
we have from (\ref{W9}):
\be
{d E\over dt} \leq C_1 E + {1\over 12 \, r_a(0)} |f(t)|^2 L^3, \label{W10}
\ee
or:
\[ E(t) \leq E(0) + C_1 \int_{0}^{t}\, E(s)\, ds + {L^3 \over 12 \, r_a(0)}
\int_{0}^{t} |f|^{2}(s')\, ds'. \]
Gronwall inequality implies:
\[ E(t) \leq \left (E(0) + {L^3 \over 12 \, r_a(0)} 
\int_{0}^{t} \, |f|^2(t') \, dt'\right )\, e^{C_1 t}, \]
or:
\be
\|(w_1,w_2)\|_{2}^{2}  \leq \min (1,m)^{-1}
\left (E(0) + {L^3 \over 12 \, r_a(0)} 
\int_{0}^{t} \, |f|^2(t') \, dt'\right )\, e^{C_1 t}. \label{W11}
\ee

Next we obtain the gradient estimates. 
Differentiating (\ref{W3})-(\ref{W4}) in $x$ gives:
\ba
& & {d\over dt} w_{1,x} =  w_{2,x} \label{W12} \\
& & {d\over dt} \left (m w_{2,x} - N \int_{0}^{x} w_2 (x',\cdot)\, dx' 
\right ) 
= \int_{0}^{x}\, \eps_s(x') \, w_2(x') \, dx' - r w_{2,x} - s' w_1 -s w_{1,x}
+ f(t) \no \\
& & -\left ( 2 r_{a}^{'} w_2 w_{2,x} + \int_{0}^{L}\, P(w_1)(x',t)\, 
K_x (x -x')\, dx' \right )\, w_2.  \label{W13}
\ea
Multiplying (\ref{W12}) and (\ref{W13}) by $w_{1,x}$ and $w_{2,x}$, and 
integrating over $x\in [0,L]$, we find:
\ba
& & {1\over 2} {d\over dt}(\|w_{1,x}\|_{2}^{2} + m \|w_{2,x}\|_{2}^{2}) 
= N {d\over dt}\int_{0}^{L}\, w_{2,x} \, \int_{0}^{x}\, w_2(x',\cdot)\, dx' + 
 (w_{1,x},w_{2,x})  \no \\
& & +\left (w_{2,x},\int_{0}^{x} \eps_{s} w_2 \right ) 
- \int_{0}^{L}\, r\, w_{2,x}^{2} 
-\int_{0}^{L} \left 
(2 r_{a}^{'} w_2 w_{2,x} + \int_{0}^{L} P(w_1) K_x \right ) w_2
w_{2,x} \no \\
& & -\int_{0}^{L}\, s' w_1 w_{2,x} - \int_{0}^{L} \, s\, w_{1,x}w_{2,x} 
+ f(t) \int_{0}^{L}\, w_{2,x}.  \label{W14}
\ea
The integral in the 
first term of the right hand side of (\ref{W14}) equals:
\be
{N\over 2} \int_{0}^{L}\, dx\, w_{2,x}(x,\cdot )\, \int_{0}^{x} 
w_2(x',\cdot )\, dx'  = -
{N\over 2} \int_{0}^{L} \, w_{2}^{2}(x',t) \, dx', \label{W15}
\ee
where we applied integration by parts once and $w_2(L,t) = 0$. 

The other terms are estimated as follows:
\ba
& & - \int_{0}^{L}\,dx\, \left (\int_{0}^{L}\, P(w_1)(x',\cdot)\, 
K_x( x - x')\, dx' 
\right )\, w_{2}\, w_{2,x} \no \\
&\leq  & C \int_{0}^{L}\, |w_{2}\,w_{2,x}|\,
 \int_{0}^{L}\, (1 + |w_1|^2)\, |K_{x}|(x - x')\, dx' \no \\
&\leq  & C (1+ \|w_1\|_{2}^{2})\int_{0}^{L}\, |w_{2}\,w_{2,x}| \no \\
& \leq & C (1+ \|w_1\|_{2}^{2})\|w_2\|_{2}\|w_{2,x}\|_{2} \no \\
& \leq & \dta_0 \|w_{2,x}\|_{2}^{2} +{1\over 4\dta_0} 
C^2(1+ \|w_1\|_{2}^{2})^{2}\|w_2\|_{2}^{2},\label{W16}
\ea
for any $\dta_0 > 0$,  $C=C(\|K_x\|_{\infty})$, by (\ref{GC}). 
Integration by parts and $w_2(L,t)=0$ give:
\be
 (w_{2,x},\int_{0}^{x} \eps_{s} w_2 ) = 
- \int_{0}^{L}\, \eps_{s}\, w_{2}^{2}\, dx \leq 0. \label{W17} 
\ee
Estimate with Cauchy-Schwarz inequalities to get:
\ba
& & - \int_{0}^{L}\, r\, w_{2,x}^{2} - 2\int_{0}^{L}\, r_{a}' w_{2}^{2} 
w_{2,x}^{2}  \leq -r_a(0) \|w_{2,x}\|_{2}^{2}; \label{W18} \\
& & -\int_{0}^{L}\, s'\, w_1 w_{2,x}\, dx - \int_{0}^{L}\, s\, w_{1,x}\, 
w_{2,x} + (w_{1,x},w_{2,x}) \no \\
&\leq & \|1-s\|_{\infty}\|w_{1,x}\|_{2}\|w_{2,x}\|_{2} + \|s'\|_{\infty} 
\|w_1 \|_{2}\|w_{2,x}\|_{2} \no \\
&\leq & 2\dta_0 \|w_{2,x}\|_{2}^{2}+ 
{1\over 16\dta_0}\|1 -s\|_{\infty}^{2}\|w_{1,x}\|_{2}^{2} 
+ {1\over 16\dta_0}\|s'\|_{\infty}^{2}\|w_1 \|_{2}^{2}; \label{W19} \\
& & f(t)\int_{0}^{L}\, w_{2,x} \leq |f(t)|L^{1/2}\|w_{2,x}\|_{2} 
\leq \dta_0 \|w_{2,x}\|_{2}^{2} +{1\over 4 \dta_0}|f(t)|^{2} L. 
\label{W20} 
\ea
Combining (\ref{W14})-(\ref{W20}), with $4\dta_0=r_{a}(0)/2$, we get:
\be
 {d\over dt} {1\over 2}(\|w_{1,x}\|_{2}^{2} + m \|w_{2,x}\|_{2}^{2}) \leq 
 -{N\over 2}{d\over dt}\, \|w_2 \|_{2}^{2}  
 -{r_a(0)\over 2}\|w_{2,x}\|_{2}^{2} + C_2(t) +
C_3(t) \|w_{1,x}\|_{2}^{2},
 \label{W20a}
\ee
where:
\ba
&&  C_{2}(t)={|f(t)|^{2} L\over 4 \dta_0} 
 + {\|s'\|_{\infty}^{2}\over 16\dta_0}\|w_1\|_{2}^{2} 
 + {C^2\over 4\dta_0}(1+ \|w_1\|_{2}^{2})^{2}\|w_2\|_{2}^{2}, \label{W20b}
\\
& & C_{3}(t)= 
{1\over 16\dta_0}\|1 -s \|_{\infty}^{2}, \label{W20c}
\ea
and $\|(w_1,w_2\|_{2}$ are bounded as in (\ref{W11}). 
Integrating (\ref{W20a}) over $t \in [0,T]$, we find:
\be
m' \|(w_{1,x},w_{2,x})\|_{2}^{2}(T) +{N\over 2}\, \|w_2\|_{2}^{2}
\leq C_{4} + \int_{0}^{T}\, C_2(t')\, dt' + 
\int_{0}^{T}\, C_3(t')\|(w_1,w_2)_{x}\|_{2}^{2}, \label{W21}
\ee
where $m' \equiv {1\over 2}\min(1,m)$, 
\be
 C_4 = {1\over 2}\max(1,m)\|(u_{0,x},u_{1,x})\|_{2}^{2} 
+ {N\over 2}\|u_1 \|_{2}^{2}; \label{W21a}
\ee
and Gronwall inequality implies:
\be
\|(w_{1,x},w_{2,x})\|_{2}^{2}(T) \leq {1\over m'} 
\left (C_4  + \int_{0}^{T}\, C_2(t)\, dt\right ) 
\exp \{{1\over m'} \int_{0}^{T}\, C_3(t)\, dt \}. \label{W22}
\ee
We see from (\ref{W4}) that $w_{2,t} \in C([0,\infty); H^1([0,L]))$, 
hence pressure $p \in C([0,\infty); H^3([0,L]))$ from (\ref{W1}).
We have thus shown:
\begin{theo}
Under the growth condition (\ref{GC}) and the initial boundary conditions 
(\ref{C4}) and (\ref{C5}), the model cochlear system (\ref{C0})-(\ref{C2}) 
has unique global solutions:
\[ (u,u_t,p) \in C([0,\infty); (H^1([0,L]))^{2}\times H^3([0,L])). \] 
\end{theo}

The estimates can be improved with the additional assumption: 
\be
s(x) \geq s_0 > 0,\;\; \forall \, x \in [0,L]; \;\; \;\; 
\|\eps_s \|_{2} < {3 r_a(0)\over 2L^{3\over 2}}. 
\label{W23}
\ee

\begin{theo}[Growth Bounds]
Under the additional assumption (\ref{W23}), the global solutions 
in Theorem 2.1 satisfy the bounds:
\ba
\|(u,u_t)\|_{2}^{2} + \int_{0}^{t}\, \|u_t \|_{2}^{2}(t')\, dt' & \leq &  
a_1 + a_2 \int_{0}^{t}\, |f(t')|^2\, dt', \no \\
\|(u_x,u_{x,t})\|_{2}^{2} + \int_{0}^{t}\, \|u_{x,t} \|_{2}^{2}(t')\, dt' 
 & \leq & a_3 + a_4 \int_{0}^{t} \,(1 + \int_{0}^{t'}\, |f(t'')|^2\, dt'')^3\, dt',
\label{gb}
\ea
for some positive constants $a_i$, $i=1,2,3,4$. 
\end{theo}

\nit {\it Proof:}
Multiplying (\ref{W3}) by $s(x) \, w_1$, 
and $(\ref{W4})$ by $w_2$, adding the two
expressions and integrating over $[0,L]$, we estimate with 
Cauchy-Schwarz inequality:
\ba
& &  (s \, w_1, w_{1,t}) + (w_2, (A w_2)_t) = 
 - \left ( \int_{x}^{L}\, dx'\, \int_{0}^{x'} \eps_{s} w_2\, dx'', w_2
\right ) \no \\ & & - (r \, w_2, w_2) + (f(t)\, (x-L), w_2 ) \no \\
& \leq & - (\int_{0}^{x}\, w_2(x')\, dx', \int_{0}^{x}\, \eps_s (x') 
w_2(x')\, dx') - r_a(0) \|w_2\|_{2}^{2} + 
f(t)(w_2,(x-L)) \no \\
& \leq & - (r_a(0) - {2\over 3} L^{3/2}\|\eps_s\|_{2})\|w_2\|_{2}^{2}
+ \dta \|w_2\|_{2}^{2} + {|f|^2\over 4\dta}\|(x-L)\|_{2}^{2}. \label{W24}
\ea
Choose $2\dta = r_a(0) - {2\over 3} L^{3/2}\|\eps_s\|_{2} > 0$ to find:
\be
{1\over 2}{d\over dt}((s\,w_1,w_1) + (Aw_2,w_2)) 
\leq  -\dta \|w_2\|_{2}^{2} + {|f|^2\over 4\dta}{L^3\over 3}. \label{W25a}
\ee
So:
\be
{1\over 2}( (s\,w_1,w_1) + (Aw_2,w_2))(t) \leq -\dta \int_{0}^{t} 
\|w_2\|_{2}^{2}\, + c_0  +
c_1 \int_{0}^{t}\, |f(t')|^2\, dt'. \label{W25}
\ee
where $c_0 = {1\over 2}( (s\,w_1,w_1) + (Aw_2,w_2))(0)$, and 
$c_1 ={L^3\over 12 \dta}$. Hence, 
\ba
{\min(s_0,m) \over 2}\|(w_1,w_2)\|_{2}^{2} & \leq & c_0 + c_1 
\int_{0}^{t}\, |f(t')|^2\, dt', \no \\
\int_{0}^{t}\, \|w_2\|_{2}^{2}(t')\, dt' & \leq & 
\dta^{-1} c_0 + \dta^{-1} \int_{0}^{t}\, |f(t')|^2\, dt'. \label{W27}
\ea  

\nit In particular, if $f(t)$ a bounded continuous function, (\ref{W27}) 
gives the growth bounds:
\be
\|(w_1, w_2) \|_{2} \leq O(t^{1/2}),\;
\int_{0}^{t}\, \|w_2\|_{2}^{2}(t')\, dt' \leq O(t), \label{W28}
\ee
implying that $\| w_2\|_{2}$ has a bounded time 
averaged $L^2$ norm square. 

Similarly, we improve the gradient estimates. Multiplying 
(\ref{W12}) by $s\, w_{1,x}$, (\ref{W13}) by $w_{2,x}$,  
integrating over $x\in [0,L]$, we cancel out the two 
integrals on $s(x)\, w_{1,x}w_{2,x}$. Proceeding as before, 
we arrive at:
\be
{d\over dt}{1\over 2}((s w_{1,x},w_{1,x}) + m\|w_{2,x}\|_{2}^{2}) 
\leq  -{N\over 2}{d\over dt} \|w_2\|_{2}^{2} 
 -{r_a(0)\over 2}\|w_{2,x}\|_{2}^{2} + C_2(t), \label{W31}
\ee
and so integrating over $[0,T]$ gives:
\ba
& & {1\over 2}((s w_{1,x},w_{1,x}) + m\|w_{2,x}\|_{2}^{2})(T) 
+ {r_a(0)\over 2}\, \int_{0}^{T}\, \|w_{2,x}\|_{2}^{2}(t')dt' + 
{N\over 2} \|w_2\|_{2}^{2}(T)  \no \\ 
& \leq & {1\over 2}((s u_{0,x},u_{0,x}) + m\|u_{1,x}\|_{2}^{2}) +{N\over 2}
\|u_1 \|_{2}^{2}
+  \int_{0}^{T}\, C_2(t')\, dt' 
\equiv C_{5,0}+ \int_{0}^{T}\, C_2(t')\, dt'. \no \\
\label{W32}
\ea
If $s(x) \geq s_0 > 0$, then:
\be
\|(w_{1,x},w_{2,x})\|_{2}^{2}(T) +{r_a(0)\over 2m''} \int_{0}^{T} 
\|w_{2,x}\|_{2}^{2}(t)\, dt \leq {1\over m''} 
\left (C_{5,0}  + \int_{0}^{T}\, C_2(t)\, dt\right ), 
\label{W29}
\ee
where $m'' ={1\over 2}\min(s_0,m)$. 
Substituting (\ref{W27}) in (\ref{W29}) gives (\ref{gb}). 
In particular, for a bounded continuous $f(t)$,
\be
\|(w_{1,x},w_{2,x})\|_{2}^{2}(T) +{1\over 2m''} \int_{0}^{T} 
\|w_{2,x}\|_{2}^{2}(t)\, dt \leq O(T^{3}). \label{W30}
\ee
The proof is finished.

If the nonlinear damping functions are bounded, i.e:
\be
r_a (\xi ) \leq C_6,\; r_a'(\xi ) \xi  \leq C_6, \; P(\xi ) \leq C_6, \; 
\forall\, \xi \geq 0, \label{ub1}
\ee
for some positive constant $C_6$, then we have:

\begin{theo}[Uniform Bounds]
Under the assumptions (\ref{W23}) and (\ref{ub1}), and that 
$f(t)$ is a bounded continuous function, the global solutions 
in Theorem 2.1 are uniformly bounded:
\[ \|(u,u_t)\|_{H^1} + \| p\|_{H^3} \leq C_7 < \infty,\;\; \forall \, t\geq 0, \]
for some positive constant $C_1$. Moreover, the dynamics admit an absorbing 
ball:
\[ \limsup_{t \to \infty} \left ( \|(u,u_t)\|_{H^1} + \| p\|_{H^3}\right ) 
\leq C_8, \]
where $C_8$ is independent of initial data. 
\end{theo}

See \cite{Strube_85} for an example of a bounded damping function. 
The energy inequality (\ref{W25a}) lacks a term like 
$-{\rm const.} \, \|w_1\|_{2}^{2}$ on the right hand side, 
and so is insufficient to provide uniform bounds. 
The idea is to bring out the skew symmetric part of the system.

\nit {\it Proof:} multiply (\ref{W3}) by $m\, w_2$, 
(\ref{W4}) by $w_1$, integrate over 
$x \in [0,L]$, and add the resulting expressions to get:
\ba
m (w_1, w_2)_{t} + (\tilde{A}w_{2,t},w_1) &=& m\|w_2\|_{2}^{2} -(s\, w_1,w_1)
- (r\, w_2,w_1) \no \\
&+ & (f(t)(x -L), w_1) - (w_1, \int_{x}^{L}\, dx'\, \int_{0}^{x'} \eps_{s} w_2\, 
d x''). \label{skw} 
\ea
Using the identity:
\[ {d\over dt}(\tilde{A}w_2,w_1)=(\tilde{A}w_{2,t},w_1)+(\tilde{A}w_2,w_{1,t})
= (\tilde{A}w_{2,t},w_1)+(\tilde{A}w_2,w_2), \]
we have:
\ba
{d\over dt}[m\, (w_1,w_2) + (\tilde{A}w_2,w_1)] & = & 
(\tilde{A}w_2,w_2) + m\|w_2\|_{2}^{2} -(s\, w_1,w_1) - (r\, w_2,w_1) \no \\
&+ & (f(t)(x -L), w_1) - (w_1, \int_{x}^{L}\, dx'\, \int_{0}^{x'} \eps_{s} w_2\, 
d x'') \no \\
& \leq & (NL^2 + m)\|w_2\|_{2}^{2} - s_0 \|w_1\|_{2}^{2} + C_6\|w_2\|_{2}\|w_1\|_{2} 
\no \\
&+ & |f|L^{3/2}\|w_1\|_{2} + L^{3/2}\, 
\|\eps_2\|_{2}\, \|w_1\|_{2}\, \|w_2\|_{2}. 
\label{W32a}
\ea
Apply Cauchy-Schwarz to polarize the last three terms to get:
\ba
C_{6}\|w_2\|_{2}\|w_1\|_{2} & \leq & {C_{6}^{2}\over s_0}\|w_2\|_{2}^{2}+
{s_{0}\over 4}\|w_1\|_{2}^{2}, \no \\  
|f|L^{3/2}\|w_1\|_{2} & \leq & |f|^2L^3/s_0 + {s_{0}\over 4}\|w_1\|_{2}^{2},\no \\
L^{3/2}\, \|\eps_2\|_{2}\, \|w_1\|_{2}\, \|w_2\|_{2} & \leq & 
{L^3 \|\eps_2\|_{2}^{2}\over s_0} \|w_2\|_{2}^{2} + {s_0\over 4} \|w_1\|_{2}^{2},
\ea
it follows from (\ref{W32a})
that:
\be
{d\over dt}[m\, (w_1,w_2) + (\tilde{A}w_2,w_1)] \leq 
C_9 \|w_2\|_{2}^{2}  - {s_0\over 4}\|w_1\|_{2}^{2} + |f|^2L^3/s_0, \label{W33}
\ee
where:
\[ C_9 = NL^2 + m + {C_{6}^{2}\over s_0} 
+ {L^3 \|\eps_2\|_{2}^{2}\over s_0}. \]

Multiplying (\ref{W25a}) by a positive constant $C_p$ and adding the resulting 
inequality to (\ref{W33}), we find:
\be
{d\over dt} E_p \leq  ( -\dta \, C_p + C_9 ) \|w_2\|_{2}^{2} 
- {s_0\over 4}\|w_1\|_{2}^{2} + |f|^2\, 
({L^3\over s_0} + {C_p L^3\over 12 \dta}), 
 \;  \label{W34}
\ee
where:
\[ E_p = {C_p\over 2}((s\,w_1,w_1) + (Aw_2,w_2))
 +m\, (w_1,w_2) + (\tilde{A}w_2,w_1). \]
 
Choose $C_p $ large enough so that $C_p > C_9 /\dta $, and:
\ba
E_p & \geq &  {C_p\over 4}((s\, w_1,w_1) + (Aw_2,w_2)) \no \\
  & \geq & 
\min(s_0,m){C_p\over 4}\|(w_1,w_2)\|_{2}^{2}; \label{W35}
\ea
so:
\[ E_p \leq {3 C_p\over 4}((s\, w_1,w_1) + (Aw_2,w_2)) \leq 
 {3 C_p\over 4}\max (\|s\|_{\infty},m +NL^2)\|(w_1,w_2)\|_{2}^{2}, \]
thanks to (\ref{W7}), then for some positive constant $C_{10}$:
\be
{d\over dt} E_p \leq - C_{10} E_p + |f|^2\, ({L^3\over s_0} 
+ {C_p L^3 \over 12 \dta}).  \;  \label{W36}
\ee

The uniform bound on $\|(u,u_t)\|_{2}$ follows from (\ref{W36}). 
Moreover, the fact that $C_{10}$ is independent of initial data implies 
the absorbing ball property of $(u,u_t)$ in $L^2$ (i.e. the limsup as 
$t \to \infty$ is bounded independent of initial data). 
Equation (\ref{W4}) and $L^{2}$ invertibility of the 
operator $A$ imply a similar uniform bound on $u_{tt}$. Equation 
(\ref{W1}) in turn shows that $\|p\|_{H^2}$ is uniformly bounded and 
has the absorbing ball property as well. 

Now we proceed with gradient estimate of $(w_1,w_2)$. The symmetric inequality
is just (\ref{W31}) but with $C_2$ uniform in time now. The skew symmetric 
inequality is obtained by multiplying $m\, w_{2,x}$ to (\ref{W12}) plus 
$w_{1,x}$ times (\ref{W13}), integrating over $x\in [0,L]$:
\ba
& & {d\over dt}[m (w_{1,x},w_{2,x})] - N (w_{1,x}, {d\over dt} 
\int_{0}^{x}\, w_{2}(x',t)\, dx') \no \\
& & \leq (w_{1,x},\int_{0}^{x}\, \eps_s \, w_{2} ) - (r w_{2,x}, w_{1,x}) 
- (s' w_1, w_{1,x}) - s_0 \|w_{1,x}\|_{2}^{2} \no \\
& & + f(t)(1, w_{1,x}) + 2 C_6 (|w_{1,x}|,|w_{2,x}|) + C_{11}(|w_2|,|w_{1,x}|). 
\label{W37}
\ea
The second term on the left hand side equals:
\[ N (w_1, w_{2,t}) = N {d\over dt}(w_{1},w_{2}) - N \|w_{2}\|_{2}^{2}. \]
It follows that:
\be
 {d\over dt}[m (w_{1,x},w_{2,x}) + N (w_1, w_2)] 
 \leq  4 C_6 \|w_{1,x}\|_{2}\|w_{2,x}\|_{2} -{s_{0}\over 2} \|w_{1,x}\|_{2}^{2}
+C_{12}, \label{W38}
\ee
for a positive constant $C_{12}$; or:
\be
 {d\over dt}[m (w_{1,x},w_{2,x}) + N (w_1, w_2)] 
 \leq  {16 C_{6}^{2}\over s_0} \|w_{2,x}\|_{2}^{2} 
-{s_{0}\over 4} \|w_{1,x}\|_{2}^{2} 
+C_{12}. \label{W39}
\ee

\nit Multiplying a constant $C'_{p} > 0$ to (\ref{W31}) with $C_2$ constant, 
and adding the resulting inequality to (\ref{W39}), we get:
\be
{d\over dt} E'_{p}  \leq  
-{s_{0}\over 4} \|w_{1,x}\|_{2}^{2} + C_{13}
 -({C'_{p}r_{a}(0)\over 2} - {16 C_{6}^{2}\over s_0})
\|w_{2,x}\|_{2}^{2},    \label{W40}
\ee
where:
\[ E'_{p} = {C'_{p}\over 2} ((s w_{1,x},w_{1,x}) + N \|w_2\|_{2}^{2}
+ m\|w_{2,x}\|_{2}^{2})
+ m (w_{1,x},w_{2,x}) + N (w_1, w_2), \]
and $C_{13} = C_{2} C'_{p} + C_{12}$. The term $(w_1,w_2)$ is bounded from 
above by constant times $\|(w_{1,x},w_{2,x})\|_{2}^{2}$ due to Poincar\'e 
inequality and so we can choose:
\[ C'_{p} > {32  C_{6}^{2}\over s_0\, r_{a}(0)}, \]
and large enough so that for some positive constants $C_{14}$, $C'_{14}$:
\be 
C'_{14}\|(w_{1,x},w_{2,x})\|_{2}^{2} \geq 
 E'_{p} \geq C_{14} \|(w_{1,x},w_{2,x})\|_{2}^{2}. \label{W41} \ee
Inequality (\ref{W40}) yields:
\be
{d\over dt} E'_{p} \leq - C_{15} E'_{p} + C_{13}, \label{W42}
\ee
implying the uniform estimate on $\|(u_{x},u_{t,x})\|_{2}$ and 
the absorbing ball property. The uniform estimate and absorbing 
ball property on $\|p\|_{H^3}$ 
follows from (\ref{W4}) and (\ref{W1}). The proof is complete.

\begin{rmk}
The estimates in Theorem 2.3 imply that the evolution map denoted by $S(t)$ 
is relatively 
compact in the space $(u,u_t)\in (L^{2}([0,L]))^{2}$. 
Hence for any bounded initial data $(u,u_t)(0) \in (H^{1}([0,L]))^2$, the 
dynamics $(u,u_t)$ approach, in the space 
$(L^{2}([0,L]))^{2}$, the universal attractor ${\bf A}$ defined as:
\[ {\bf A} = \cap_{t > 0} S(t) B_{\rho_0}, \]
where $B_{\rho_0}$ denotes the ball of radius $\rho_0$ in $(H^{1}([0,L]))^{2}$, 
the absorbing ball given by the estimates of Theorem 2.3. 
\end{rmk}

\section{Multitone Solutions}
\setcounter{equation}{0}
In this section, we consider special solutions to (\ref{C0})-(\ref{C1}) 
that exhibit explicitly their frequency contents. 
For simplicity, 
let us assume that $\eps_{s}(x) = 0$, 
and $s(x) \geq s_0 > 0$, $\forall \, x \in [0,L]$.
\subsection{Linear Waves}
First we consider the linear regime with $r= r_0$,  a positive constant. 
Solutions are superpositions of single frequency time harmonic waves 
of the form $p = P(x)e^{i\omega t} + c.c.$, $u=U(x)e^{i\omega t} + c.c$, 
$c.c$ complex conjugate, where $P$ and $U$ are complex functions that satisfy: 
\ba
P = ( - m\, \om^2 + i \, r_0\, \om + s(x) )\, U, \label{T1} \\
P_{xx} + N\, \om^2 \, U = 0, \label{T2} \\
P_{x}(0)= P_{in}, \; P(L)=0. \label{T3}
\ea
Let:
\[ (\al + i \beta )(x) = { N \, \om^2 \over 
-m\om^2 + i r_0 \om  + s(x) }, \]
so:
\be
(\al,\beta) = N\, \omega^2\, ( -m\, \omega^2 + s(x), -r_0\, \omega)/ 
[ (-m\, \omega^2 +s(x))^{2} + r_{0}^{2}\, \omega^2 ], \label{T4}
\ee 
then (\ref{T1})-(\ref{T3}) is equivalent to:
\be
P_{xx} + (\al (x) + i \beta (x)) \, P = 0, \label{T5}
\ee
subject to (\ref{T3}). If $\omega \not = 0$, $\beta \not =0$. 
We show:
\begin{lem} 
The boundary value problem, (\ref{T5}) and (\ref{T3}), has a unique 
solution for all $\omega $ such that $\|P\|_{H^2([0,L])}\leq C |P_{in}|$, 
for some constant $C$ independent of $\omega$. 
\end{lem}
Proof: Write $P = P_{in}(x - L) + Q$, then $Q$ satisfies:
\be
Q_{xx} + (\al (x) + i \beta (x)) \, Q = -(\al (x) + i\beta (x)) (x -L) P_{in}
\equiv (f_1 + i f_2)(x), 
\label{T6}
\ee
with boundary conditions $Q_{x}(0)=0$, $Q(L)=0$. 
The left hand side is a Fredholm operator on $Q$, so it is sufficient 
to prove that zero is not an eigenvalue, which follows from the 
estimate below. Write $Q = q_1 + i q_2$, then:
\be
\left ( \begin{array}{r}
q_{1} \\
q_{2}
\end{array} \right )_{xx} + 
\left (\begin{array}{rr}
\alpha & -\beta  \\
\beta  & \alpha 
\end{array} \right ) 
\left (\begin{array}{r}
q_{1} \\
q_{2}
\end{array} \right ) = \left ( \begin{array}{r}
f_1 \\
f_2
\end{array} \right ). \label{T7}
\ee

Multiplying (\ref{T7}) by $(q_2, -q_1)$, we find:
\[ q_{1,xx}q_2 - q_1 q_{2,xx} - \beta q_{2}^{2} -\beta q_{1}^{2} = 
f_1 q_2 - f_2 q_1, \]
which gives upon integrating over $x \in [0,L]$, integrating by parts:
\be 
-\int_{0}^{L}\, \beta \, (q_{1}^{2} +q_{2}^{2})\, dx = 
\int_{0}^{L} (f_1 q_2 - f_2 q_1)\, dx. \label{T8}
\ee
Multiplying (\ref{T7}) by $(q_1,q_2)$, integrating over 
$x \in [0,L]$, we have after integration by parts:
\be
 -\| (q_{1,x}, q_{2,x})\|_{2}^{2} + \int_{0}^{L}\, \alpha \, 
(q_{1}^{2} + q_{2}^{2})\, = \int_{0}^{L}\, (f_1 q_1 + f_2 q_2). \label{T9}
\ee
It follows from (\ref{T9}) and Poincar\'e inequality that:
\ba
 \|(q_1,q_2)\|_{2}^{2} & \leq & L^2 \|(q_{1,x},q_{2,x})\|_{2}^{2} \no \\
& \leq & L^2 \|\alpha \|_{\infty} \, \|(q_1,q_2)\|_{2}^{2} 
+ L^2 \|(f_1,f_2)\|_{2} \, \|(q_1,q_2)\|_{2}.
\label{T10}
\ea
For $|\omega |\leq \omega_0 \ll 1$, 
$\|\alpha \|_{\infty} \leq 2N\omega^2/s_0$,
$2 \omega^2 L^2 N/s_0 \leq 1/2$, 
\be
\|(q_1,q_2)\|_{2} \leq 2L^{2} \|(f_1,f_2)\|_{2}. \label{T11}
\ee
As $\|(f_1,f_2)\|_{2} = O(\omega^2)$, (\ref{T11}) implies 
$ \|(q_1,q_2)\|_{2} = O(\omega^2)$ for $\omega \ll 1$. 
For $\omega^{2} \geq M=M(N,m,s,r_0)$, $M$ large enough, 
$\al \sim - {N\over m}$, (\ref{T9}) shows:
\be \|(q_1,q_2)\|_{2} \leq {2 m\over N} \|(f_1,f_2)\|_{2}. \label{T11a}
\ee

When $\omega_{0}^{2} \leq \omega^2 \leq M$, 
$|\beta |$ is bounded from below 
uniformly in $\omega $:
\[ |\beta | \geq {N\, r_0 \, |\omega|^3\over \|-m\omega^2 + s(x)\|_{\infty}^{2}
+ r_{0}^{2}\omega^2 }\geq \beta_0, \]
for some positive constant $\beta_0$ only 
depending on $r_0$, $m$, $L$, and $s(x)$. 
Inequality (\ref{T8}) gives:
\be
\|(q_1,q_2)\|_{2} \leq \beta_{0}^{-1} \|(f_1,f_2)\|_{2} 
= \beta_{0}^{-1} C_1(N,m,r_0,L,s) |P_{in}|, \label{T12}
\ee
for positive constant $C_1(N,m,r_0,L,s)$, 
uniformly in $\omega^2 \in [\omega_{0}^{2}, M]$.
Combining (\ref{T11}), (\ref{T11a}), and (\ref{T12}), we see that  
for any $\omega$, and any given $P_{in}$, there is a unique solution $P$, 
$\|P\|_{2} \leq C_2 |P_{in}|$, for constant $C_2$ independent of $\omega$.
The lemma is proved by applying the $L^2$ estimate and the $P$ equation 
(\ref{T5}). 

\subsection{Nonlinear Waves}
We are interested in the persistence of multitone solutions 
when nonlinearities are present. For simplicity, we shall consider: (A1) 
$r_a > 0$, a constant, and $P(u)=u^2$, the overall nonlinearity 
is cubic. As for linear waves, assume that 
(A2) $\eps_s = 0$, $s(x) \geq s_0 > 0$, $s \in C^1([0,L])$. We prove:
\begin{theo}[Existence and Uniqueness of Multitone Solutions]
Let the left boundary condition be:
\[ f_{in}(t) = \sum_{j=1,\cdots,m}\, a_j\, \exp\{i\omega_j t \} + c.c., \]
and fix $\rho \geq 1$. Then under (A1-(A2) and 
for $\gamma $ small enough (independent of $\rho $), 
system (\ref{C0})-(\ref{C1}) has a unique solution of the form:
\be
u(x,t)=\sum_{k \in Z^m}\, U_{k}(x)\, \exp\{ik\cdot \omega t \} + c.c, 
\; \label{mt}
\ee
where $\omega = (\omega_1,\omega_2,\cdots,\omega_m)$, 
and complex valued functions $U_k(x) \in H^{1}([0,L])$, 
such that:
\be
\|u \| \equiv \sum_{k}\, \rho^{|k|}\, \|U_{k}\|_{H^1} < \infty. \label{T13}
\ee
The pressure $p$ is similar.
\end{theo}

\nit {\it Proof:} Let $B$ be the Banach space consisting of space-time 
functions of the form (\ref{mt}) with norm (\ref{T13}). Let $B_1 = \{ 
v \in B: v_t \in B \}$, a subspace of $B$.
Consider the mapping $M: v \in B_1 \ra u$ defined as the unique 
bounded solution of 
the following equation in $B_1$:
\ba
m\, u_{tt} & + & \int_{x}^{L}\, dx'\, \int_{0}^{x'}\, dx''\, N u_{tt} + 
f_{in}(t)(L -x)  \no \\
& & + r_a\,  u_{t} + s(x) \, u = -\gam \, r_{nl}(x,v^2)\, v_t, \label{T14}
\ea
where $\gamma \, r_{nl}$ is the nonlinear nonlocal part of the damping 
function. 

Let us show that $M$ is a well-defined bounded 
mapping from $B_1$ to itself. 
First we notice that for any functions $u_i \in B$, $i=1,2$:
\ba
u_1\cdot u_2 & = & \sum_{k_1\in Z^m}\, u_{1,k_1}\, e^{ik_1\cdot \omega \,t }
\sum_{k_2 \in Z^m}\, u_{2,k_2}\, e^{ik_2 \cdot \omega \, t} \no \\
& = & \sum_{j\in Z^m} (\sum_{k_1 +k_2 = j}\, u_{1,k_1}\, u_{2,k_2}\, )
e^{i j\cdot \omega \, t}, \no
\ea
so:
\ba
\|u_1 u_2\|& = & \sum_{j \in Z^m}\, \rho^{|j|}\, 
\|\sum_{k_1 + k_2 = j} u_{k_1} u_{k_2} \|_{H^1} \no \\
& & \leq \sum_{k_1,k_2 \in Z^m}\, \rho^{|k_1| + |k_2|}\|u_{k_1} u_{k_2} \|_{H^1} 
\no \\
& & = \sum_{k_1 \in Z^m} \rho^{|k_1|} \|u_{k_1}\|_{H^1}
\sum_{k_2 \in Z^m} \rho^{|k_2|} \|u_{k_2}\|_{H^1} =\|u_1\|\|u_2\|. \label{T15}
\ea
It follows that:
\be
\| r_{nl}(x,v^2)v_t \| \leq C \| v^2*K \| \|v_t\| \leq C' \|v^2\| \|v_t\| 
= C' \|v\|^{2}\, \|v_t\|, \label{T16}
\ee
where $C'$ depends on the kernel function $K$, and $*$ denotes the convolution 
integral on $[0,L]$. 

\nit Denoting $F(x,t) = -\gam \, r_{nl}(x,v^2)\, v_t \in B$, we show that 
$u \in B_1$. Write $F(x,t)=\sum_{k\in Z^m} F_k \, e^{ik\cdot \omega t}$ + c.c, 
then 
(\ref{T14}) is same as the system:
\ba
p_{xx} & - & N u_{tt} = 0, \no \\
p & = & m u_{tt} + r_a\, u_t + s(x) u + F, \label{T16a}
\ea
with boundary condition: $p_x(0,t)=f_{in}(t)$, $p(L,t)=0$. 
Seek solution of system (\ref{T16a}) in the form:
\[ p =\sum_{k \in Z^m}\, p_{k}e^{ik \cdot \omega t} + c.c, \]
\[ u = \sum_{k \in Z^m}\, u_k e^{ik \cdot \omega t} + c.c, \]
resulting in ($k=(k_1,k_2,\cdots, k_m)$):
\ba
p_{k} = ( -m (k \cdot \omega )^2 + i r_a (k\cdot \omega ) + s(x))u_{k} + F_k, 
\no \\
p_{k,xx} + N(k\cdot \omega )^2 u_{k} =0, \label{T17}
\ea
with boundary condition: $p_{k}(L)=0$, 
$p_{k,x}(0) = 0$ if $k$ is not one of the $m$ modes
(along $e_j$, $j =1,2,\cdots, m$) of $f_{in}$; otherwise $p_{k,x}(0)= 
a_{j}$, if $k = e_j$. 

\nit For each $k$, the system (\ref{T17}) can be uniquely solved as in 
Lemma 4.1, with the estimates:
\ba
\|p_{k}\|_{H^1} & \leq & C_1 \|F_{k}\|_{2} + C_2 \|f_{in,k}\|_{2}, \no \\
\|u_{k}\|_{H^1} & \leq & C_3 \|p_{k}\|_{H^1}/(1 + |k\cdot \omega |). \label{T18}
\ea
It follows that the mapping $M$ is from $B_1$ to itself, and it is 
not hard to check that $M$ 
is a contraction mapping if $\gamma $ is small enough. The proof is done.


\section{Numerical Results}
\setcounter{equation}{0}
The model system is computed with a second order semi-implicit finite 
difference method, we refer to \cite{Xin_02} for its details and 
choice of the coefficient functions in the model other than $P(u)=u^2$. 
The input at the left boundary $x=0$ 
is the sum of two tones (sinusoids) 
at frequencies $f_1 = 3.5$ kHz (kilo Hertz) and $f_2 =4 $ kHz, with 
amplitudes 80 dB (decibel) and 85 dB respectively. The zero decible is 
20 $\mu $ Pa in physical unit. 
The time step is 0.01 ms (millisecond), and spatial grid is 0.01 cm. 
The computation ends at 20 ms when the BM responses reach a steadily 
oscillating state. To observe the frequency content of such a state, 
we select four points ($x_j$'s, $j=1,2,3,4$) on BM, and 
examine the response time series ($u(x_j,t)$) at these 
points from 5 ms to 20 ms (to omit initial transient effects). 
The power spectral density of $u(x_j,t)$ at each $j$ is obtained 
using signal processing tool (sptool) in Matlab, to illustrate the 
energy distribution across frequencies in log-scale.  

\begin{figure}[p]
\centerline{\includegraphics[width=400pt,height=150pt]{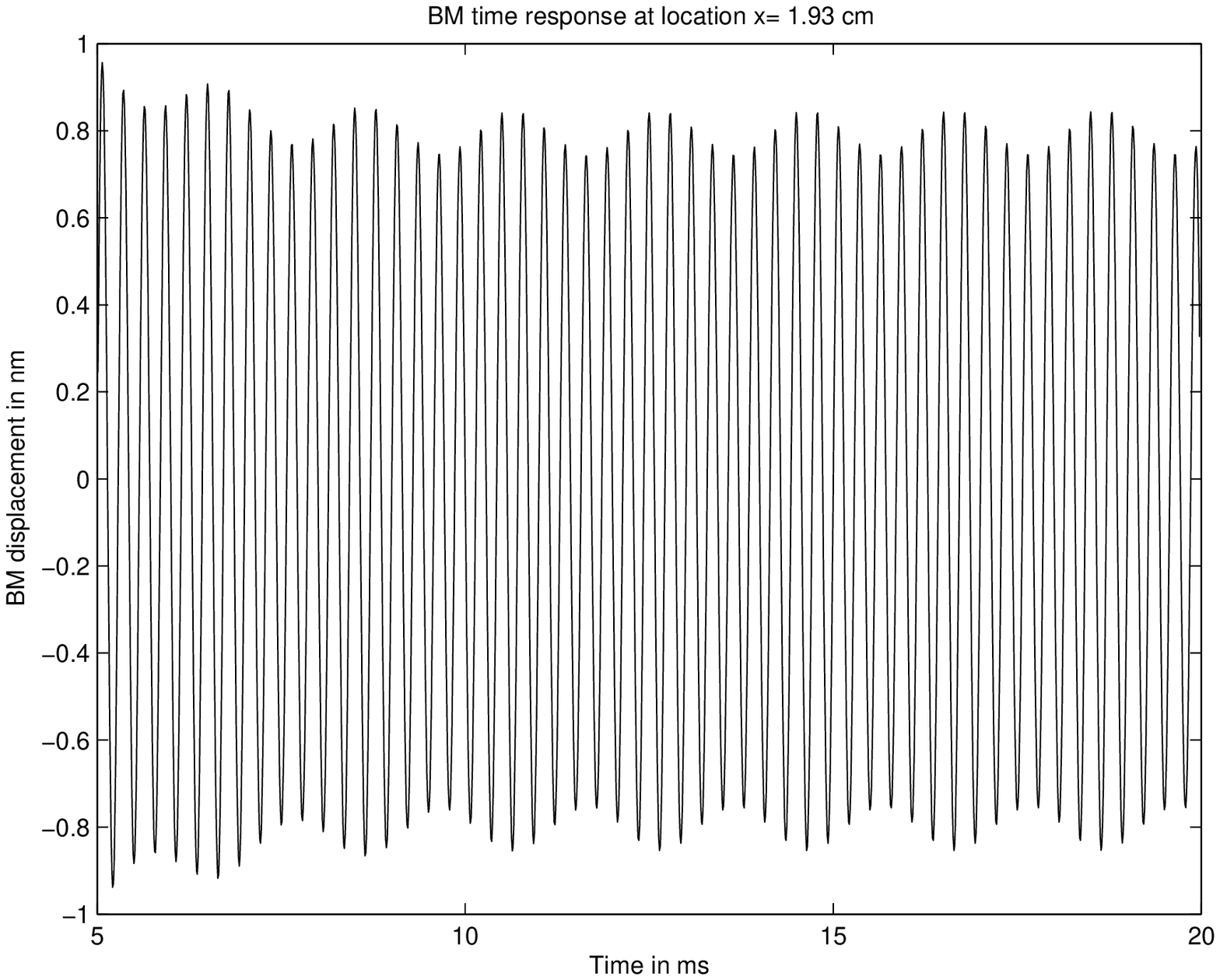}}
\bigskip

\centerline{\includegraphics[width=400pt,height=350pt]{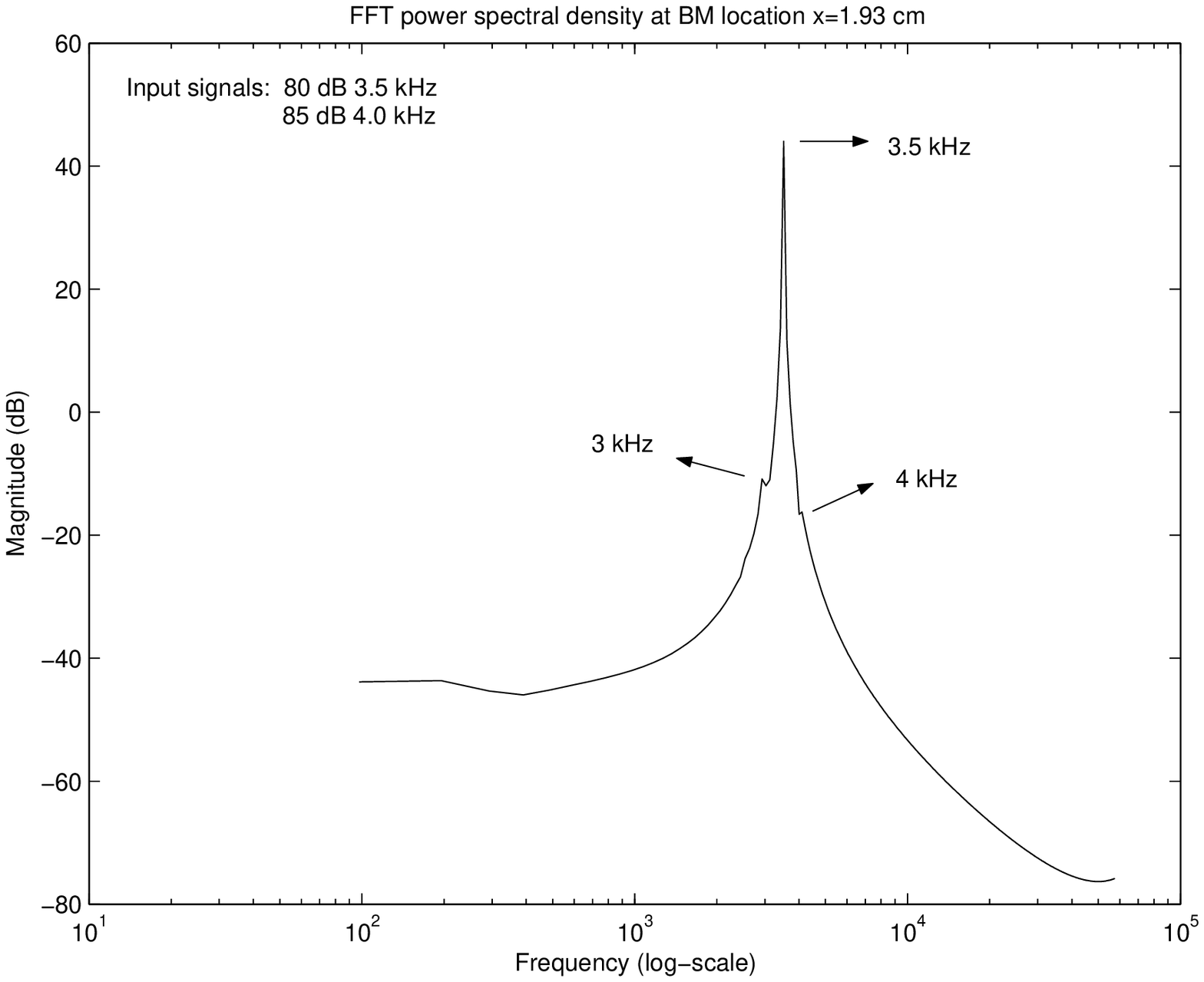}}
\caption{Time series of BM displacement at $x=1.93 $ cm (top frame),
and its FFT power spectral density vs. frequency (bottom frame).}
\label{Fig1}
\end{figure}

\begin{figure}[p]
\centerline{\includegraphics[width=400pt,height=150pt]{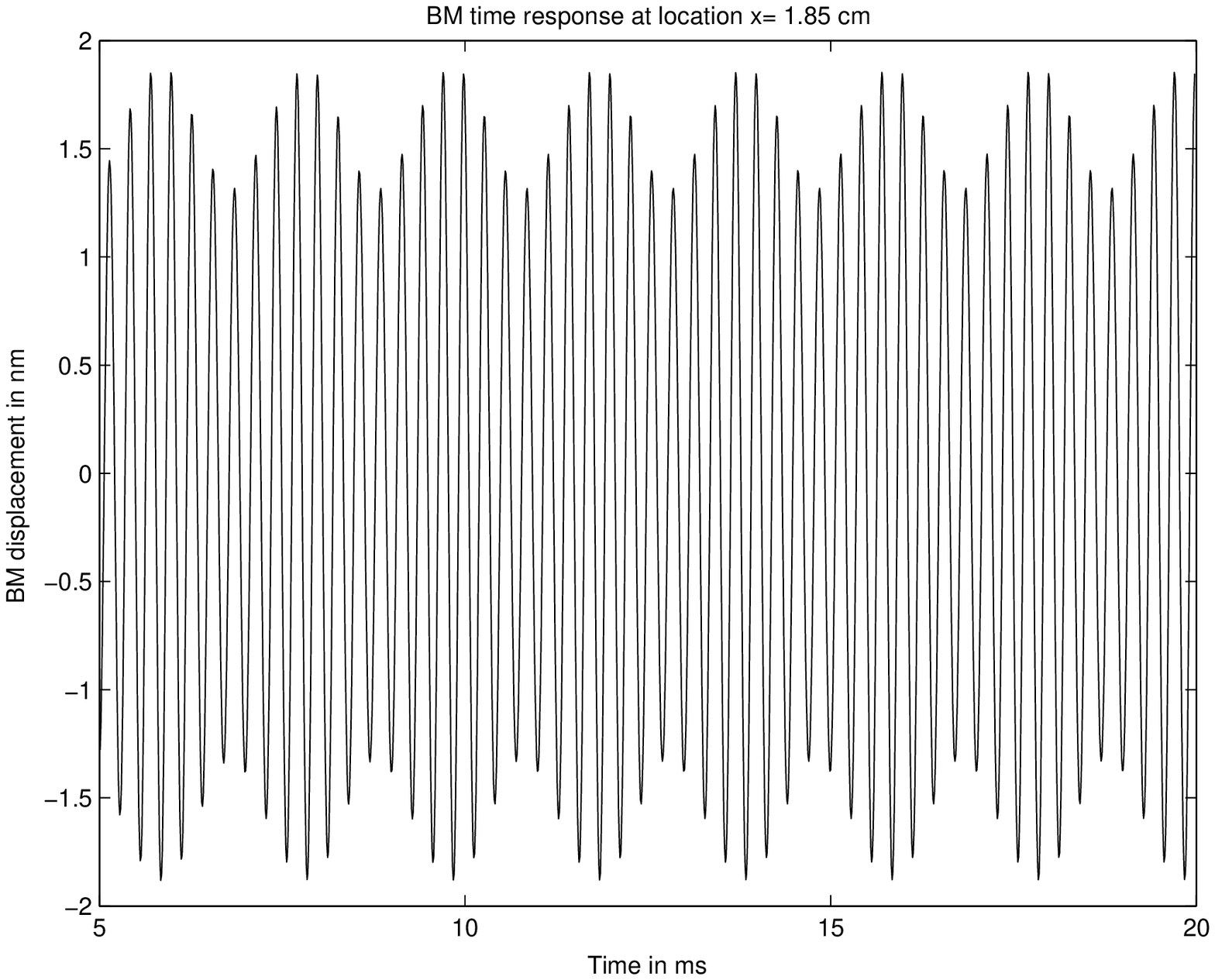}}
\bigskip

\centerline{\includegraphics[width=400pt,height=350pt]{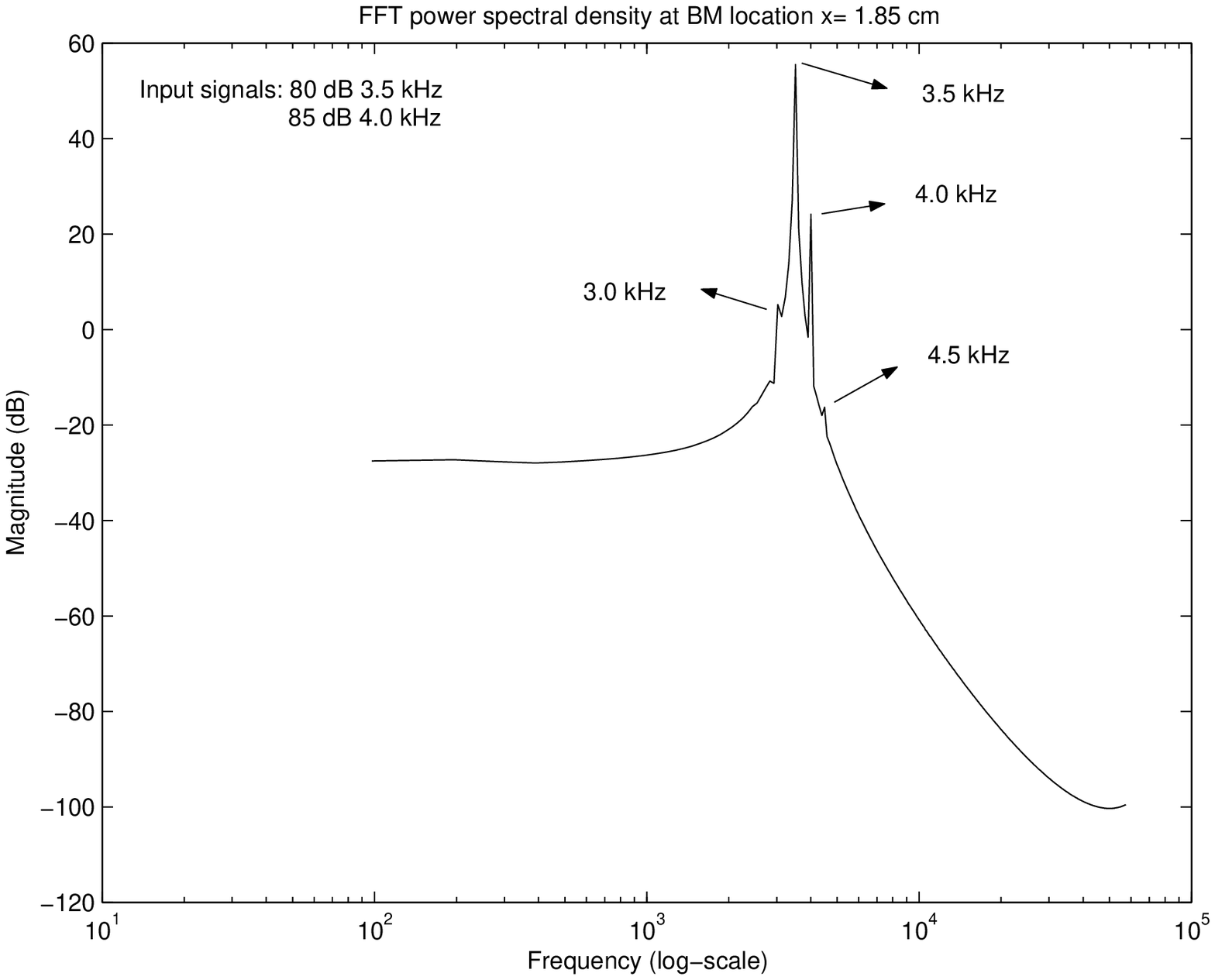}}
\caption{Time series of BM displacement at $x=1.85 $ cm (top frame),
and its FFT power spectral density vs. frequency (bottom frame).}
\label{Fig2}
\end{figure}

\begin{figure}[p]
\centerline{\includegraphics[width=400pt,height=150pt]{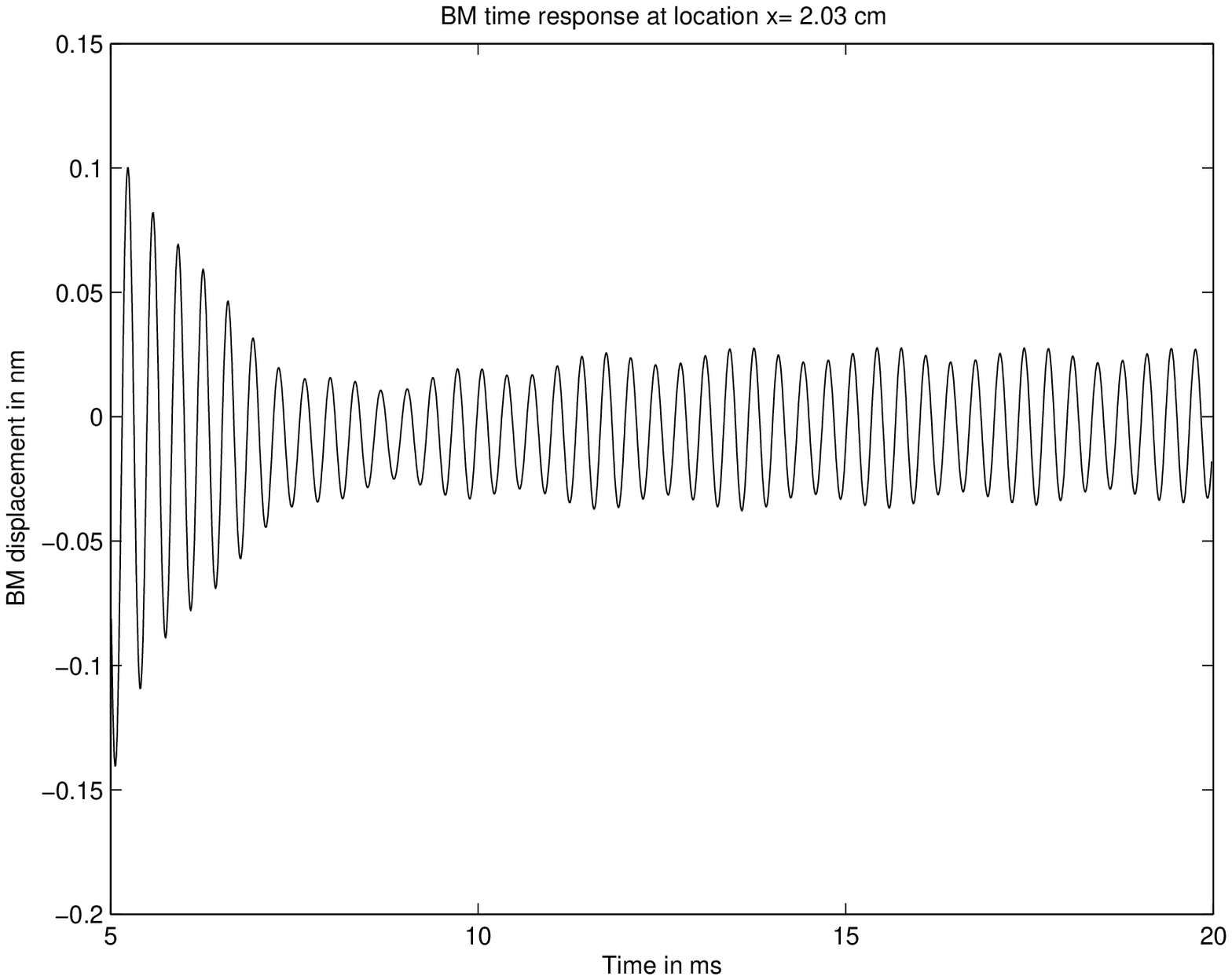}}
\bigskip

\centerline{\includegraphics[width=400pt,height=350pt]{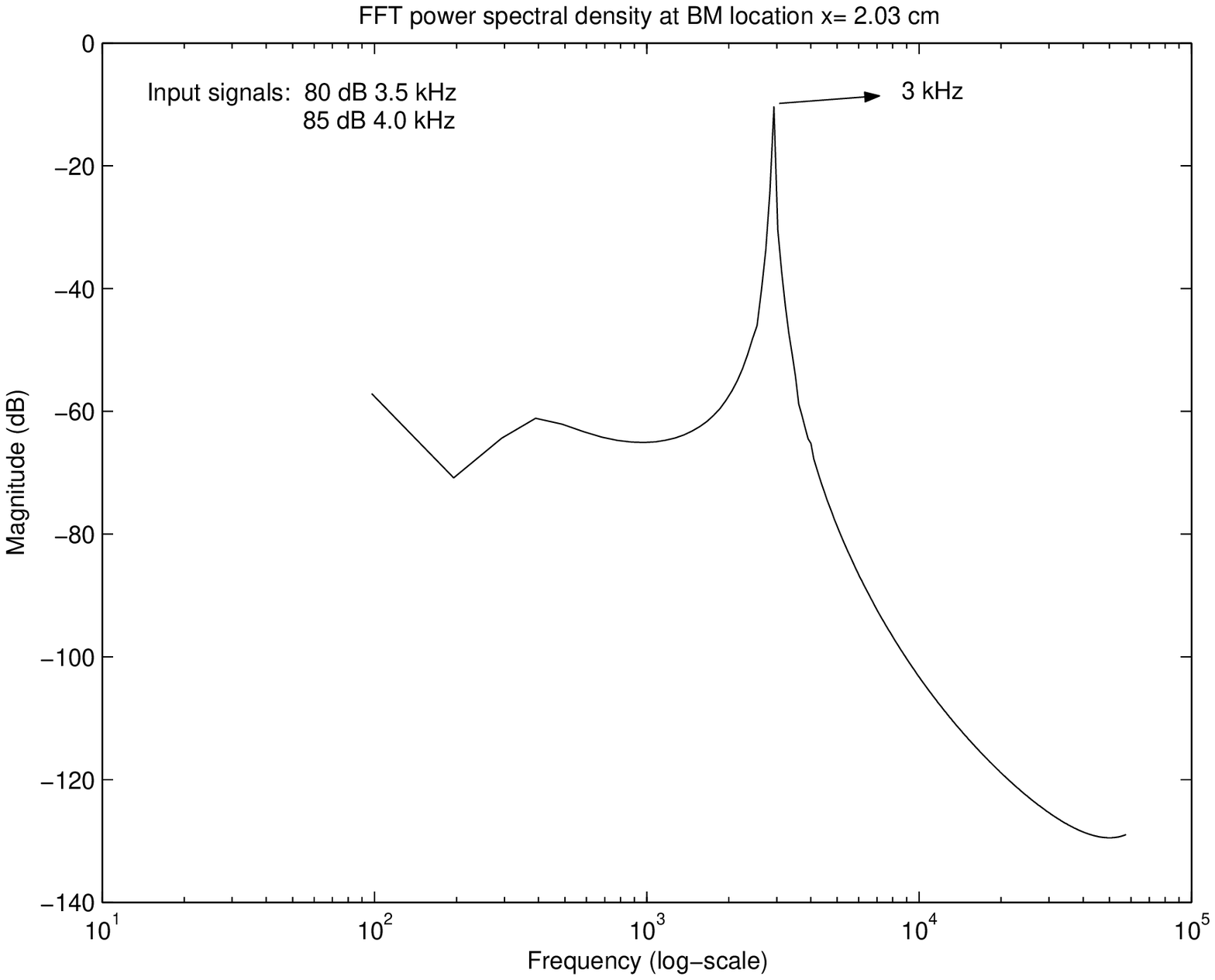}}
\caption{Time series of BM displacement at $x=2.03 $ cm (top frame),
and its FFT power spectral density vs. frequency (bottom frame).}
\label{Fig3}
\end{figure}

\begin{figure}[p]
\centerline{\includegraphics[width=400pt,height=150pt]{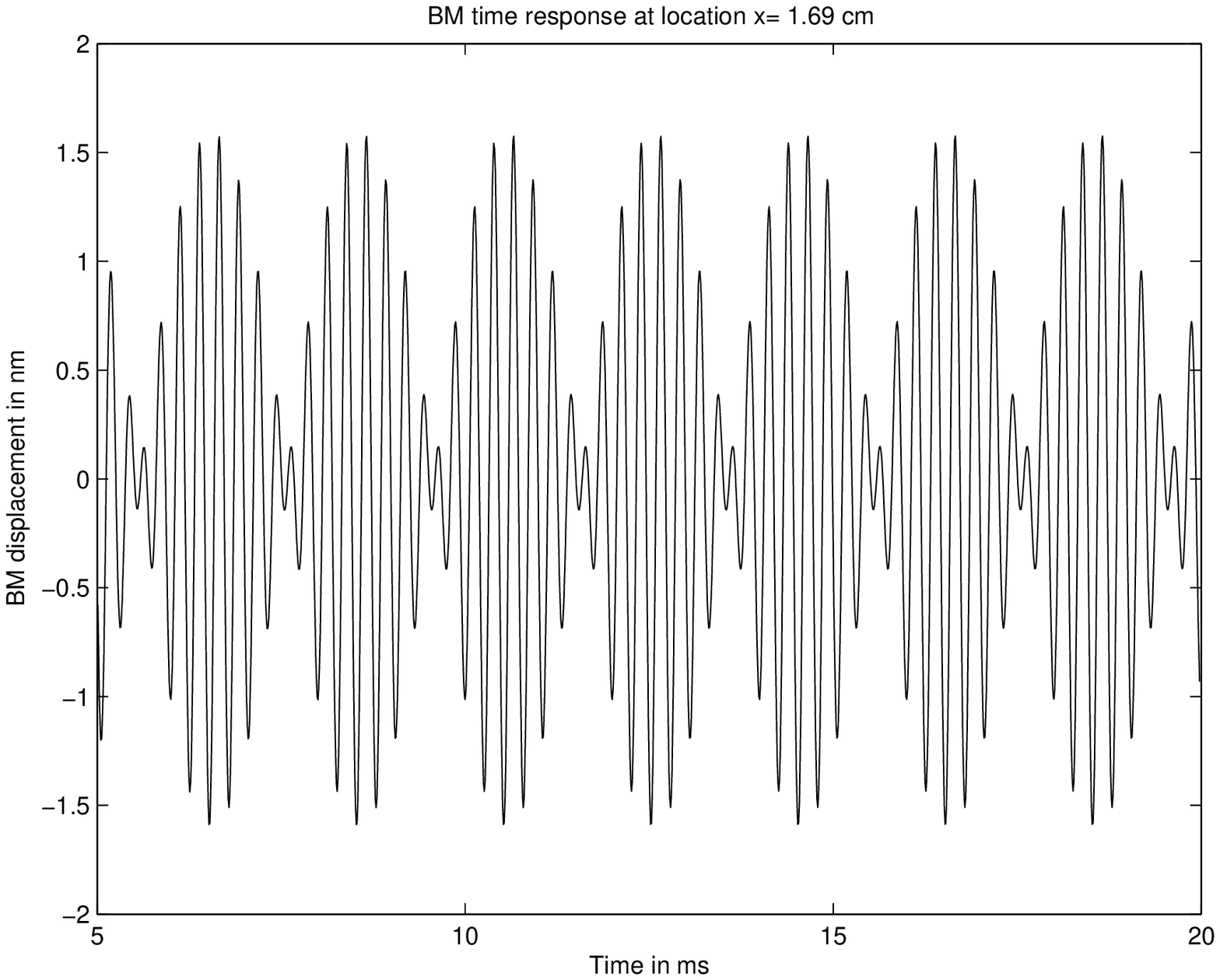}}
\bigskip

\centerline{\includegraphics[width=400pt,height=350pt]{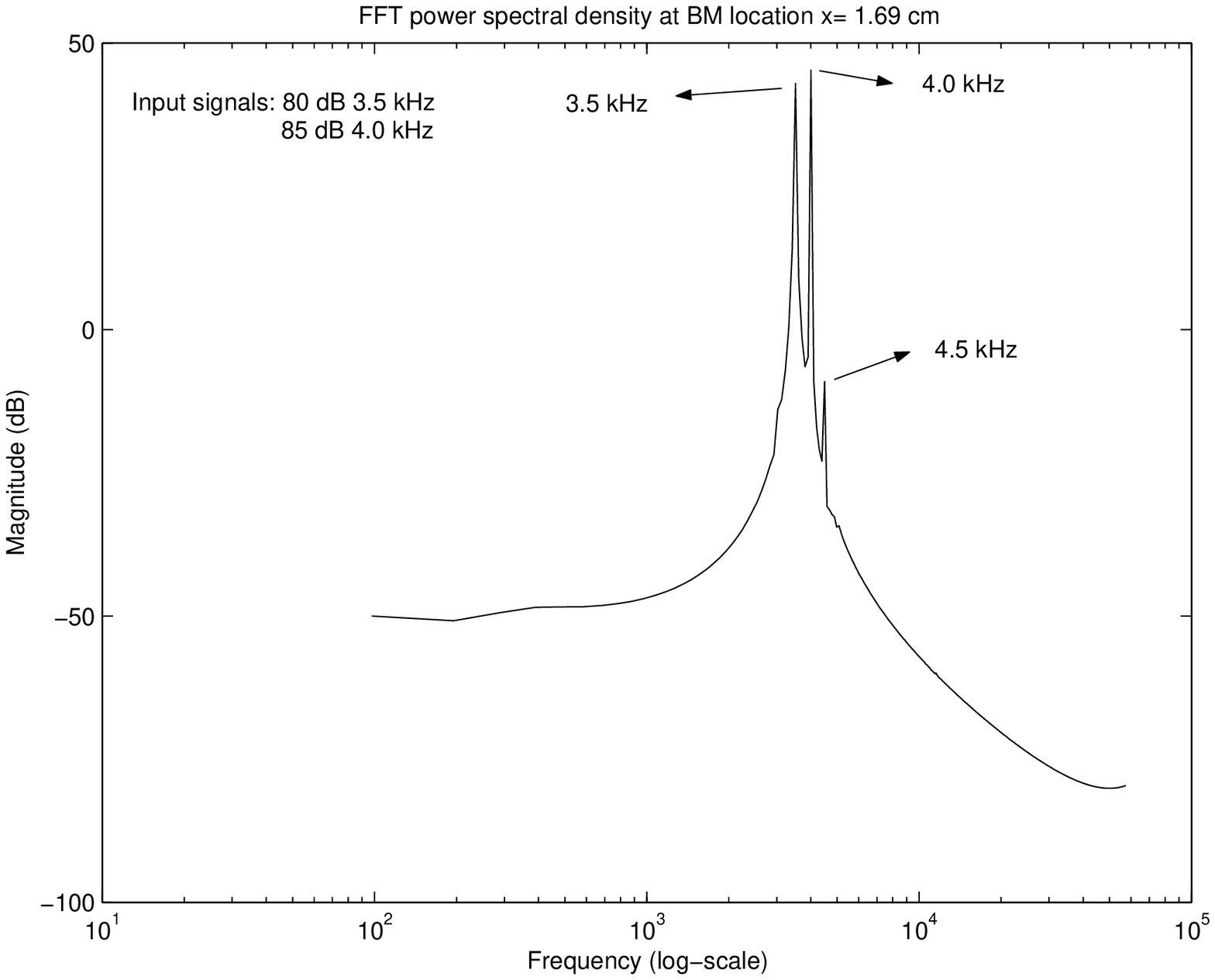}}
\caption{Time series of BM displacement at $x=1.69 $ cm (top frame),
and its FFT power spectral density vs. frequency (bottom frame).}
\label{Fig4}
\end{figure}

Figure (\ref{Fig1}) top frame 
shows the time series of BM displacement at $x =1.93 $ cm, 
the bottom frame is the log-log plot of 
FFT power spectral density vs.  
frequency. We see the major peak at 3.5 kHz, as $x=1.93$ cm is 
the (so called characteristic) location for the peak of a single 
3.5 kHz tone. In addition, we see 
two small side 
peaks at 3 kHz (= $2f_1 -f_2$), and 4 kHz ($f_2 $). 
In figure (\ref{Fig2}), at $x=1.85 $ cm, the characteristic location 
for $f_2$, the $f_2$ peak is more pronounced, however the $f_1$ peak is 
still the highest. Such effect of lower frequency tone $f_1$ 
to a higher frequency 
tone $f_2$ is called upward masking in hearing. In addition, there are 
two small side peaks at 3.0 kHz ( $=2f_1 -f_2$ ) and 4.5 kHz 
($= 2f_2 - f_1 $). In figure (\ref{Fig3}), at $x=2.03 $, 
the characteristic location for 3 kHz, a dominant single peak 
due to the generated combination tone $2f_1 -f_2 $ is observed.
In contrast, the $2f_2 -f_1 $ tone (= 4.5 kHz) is weaker and  
dominated by $f_1$ and $f_2$  
even at its characteristic location $x=1.69$ cm, see 
figure (\ref{Fig4}). 
The above findings on combination tones are consistent with the 
experiments on cochlea \cite{RR_01}, and the 
analytic structures of multi-tone solutions in the previous section.  

%
%
%
%
%
%
%
%
%
%
%

\section{Conclusions}
The nonlinear nonlocal cochlear models of the transmission line type 
are well-posed globally in 
time and admit exact multi-frequency solutions in the weakly 
cubic nonlinear regime. For finitely many tonal input at distinct 
frequencies, the exact solutions contain all integral linear combinations 
of input frequencies. For two tone input with frequencies 
$f_1$ and $f_2$ at high enough intensities, we observed numerically 
the combination tones $2 f_1 -f_2$ and $2 f_2 -f_1$ in model output, 
in agreement with existing experimental observations \cite{RR_01} and 
the structure of analytical solutions. 

\section{Acknowledgements}
J. X. would like to thank P. Collet for a stimulating 
discussion of quasiperiodic solutions in dissipative systems, and a 
visiting professorship at the 
Inst. H. Poincar\'e, where the work was in progress. 
He also thanks H. Berestycki, P. Constantin, G. Papanicolaou, and 
J-M Roquejoffre for their interest.   

\nit Part of the work was done while Y.Q. was a TICAM (Texas Institute of 
Computational and Applied Math) visiting fellow 
at the University of Texas at Austin. The TICAM research fellowship 
is gratefully acknowledged. 

\bibliographystyle{plain}

\newpage

\newpage

\newpage

\end{document}